\documentclass[12pt]{article}
\usepackage[super,compress]{cite}
\usepackage{graphicx}
\usepackage{epsfig}

\newcommand{\be}{\begin{equation}}
\newcommand{\ee}{\end{equation}}

\newcommand{\bea}{\begin{eqnarray}}
\newcommand{\eea}{\end{eqnarray}}
\newcommand{\beq}{\begin{equation}}
\newcommand{\eeq}{\end{equation}}
\newcommand{\nn}{\nonumber}

\def\fun#1#2{\lower3.6pt\vbox{\baselineskip0pt\lineskip.9pt
\ialign{$\mathsurround=0pt#1\hfil##\hfil$\crcr#2\crcr\sim\crcr}}}

\begin{document}

\title{Narrow pentaquarks as diquark-diquark-antiquark systems}

\author{V.V. Anisovich$^{+}$, M.A. Matveev$^+$, J. Nyiri$^*$,  A.N. Semenova$^+$,
}
\maketitle

\begin{center}
{\it $^+$Petersburg Nuclear Physics Institute of National Research Centre ''Kurchatov Institute'', Gatchina, 188300, Russia}

{\it $^*$Institute for Particle and Nuclear Physics, Wigner RCP,
Budapest 1121, Hungary}

\end{center}

\begin{abstract}
The diquark-diquark-antiquark model describes pentaquark states both in terms of quarks and hadrons. The latest LHCb data for pentaquarks with open charm emphasize the importance of hadron components in the structure of pentaquarks. We discuss pentaquark states with hidden charm $P(\bar c cuud)$ and those with open charm $P(\bar uussc)$ which were discovered recently in LHCb data ($J/\Psi p$ and $\Xi_c^+ K^-$ spectra correspondingly). Considering the observed states as members of the lowest ($s$-wave) multiplet, we discuss the mass splitting of states and the dumping of their widths.
\end{abstract}

Keywords: Quark model; resonance; exotic states.

PACS numbers: 12.40.Yx, 12.39.Mk, 14.20.Lq

\section{Introduction}

A narrow peak in $J/\Psi p$ spectra was seen in LHCb data\cite{LHCb} and was interpreted as a pentaquark. In Refs.~[\citen{maia-polo,ala-2,ccstr,wang}] it was considered as an antiquark-diquark-diquark system $P=\bar c\cdot (cu)\cdot (ud)$.

The recent discovery of the system of resonances which can be interpreted as a pentaquark with open charm attracts again the attention to the problem of many-quark systems. Five narrow peaks which are seen in the $\Xi_c^+ K^-$  channel\cite{LHCbnew} can be interpreted as $P =\bar u\cdot (cs)\cdot (su)$ and $P =\bar u\cdot (cu)\cdot (ss)$ states.

However, in the channel $\Xi_c^+ K^+$ analogous narrow states are not observed. In the quark language these two channels differ in the permutation of $u$ and $s$ quarks: $\bar u\cdot (cs)\cdot (su)\to\bar s\cdot (cu)\cdot (su)$ and $\bar u\cdot (cu)\cdot (ss)\to\bar s\cdot (cs)\cdot (uu)$. Nevertheless, from the point of view of the quark composition, the absence of states in the second channel may be surprising. In the terms of the hadronic component it is not so.

The hadronic component was analyzed for tetraquarks.\cite{QQQQ} This component is formed by quark recombination of the type $(q_1q_2)\cdot(\bar q_3\bar q_4)\to(q_1\bar q_3)\cdot(q_2\bar q_4)+(q_1\bar q_4)\cdot(q_2\bar q_3)$. It is seen that the same content of quarks can produce different hadrons with correspondingly different threshold singularities. Different probabilities of transition of exotic states to hadrons are essential for their mass shifts. The effects of influence of hadronic components on pentaquark masses are crucial for their observation. In the present paper we discuss shortly the mass shifts for both pentaquark with hidden charm and pentaquark with open charm in the framework of diquark-diquark-antiquark approach; a more complete consideration of the hadronic components for pentaquarks is discussed in the subsequent papers.

\section{Hidden charm pentaquarks $P(\bar ccq_1q_2q_3 )$}\label{sec1}

A pentaquark states with hidden charm $P(\bar c cuud)=P(\bar c cqq'q'')$ were observed by the LHCb collaboration.\cite{LHCb} There were detected broad state $P(4380)$ and narrow one $P(4450)$. These resonances generated a wide discussion.\cite{wang,Petrov,espo,os,pol,ali,skw,wo,st}. We will concentrate on discussing the narrow state because there are different variants of interpretation for broad state (for example see Ref.~[\citen{odnakartinka}]). In Refs.~[\citen{ala-2,ccstr}] it was proposed that the lowest pentaquark states are $s$-wave systems with $P(4450)$ being a $\frac{5}{2}^-$ state.

It is reasonable to discuss mass splitting of pentaquarks on the basis of results for standard mesons $(q\bar q)$ and baryons $(qqq)$. It was understood relatively long ago that the mass splitting of hadrons can be well described in the framework of the quark model by the short-ranged color-magnetic interactions of the constituents.\cite{ZSah,ruh-gl,GL} For mesons and baryons the mass formulae discussed by Glashow are:\cite{GL}
\bea \label{7}
&&
M_M=\sum\limits_{j=1,2}m_{q(j)}+a
\frac{\vec{s}_1\vec{s}_2}{m_{q(1)}m_{q(2)}},
\\
&&
M_B=\sum\limits_{j=1,2,3}m_{q(j)}+b
\sum\limits_{j>\ell}\frac{\vec{s}_j\vec{s}_\ell}{m_{q(j)}m_{q(\ell)}},
\nn
\eea
where  $\vec{s}_j$ and $m_{q(j)}$ refer to spins and masses of the constituents,$\Delta$ is the parameter of spin splitting. Mass splitting parameters in eq.~(\ref{7}), $a$ and $b$, are characterized by the size of the color-magnetic interaction and the size of the discussed hadron, the short-range interaction is supposed in Ref.~[\citen{GL}]. For the 36-plet mesons ($q\bar q$) and 56-plet baryons ($qqq$) formulae of eq. (\ref{7}) work well.

It seems natural to apply modified formulae (\ref{7}) to pentaquark systems. For multiquark states from Ref.~[\citen{LHCb}] we have:
\bea \label{8}&&
M_{q_1q_2\cdot q_3c\cdot \bar c}=3m_{D(q_1q_2)D(q_3c)\bar c}
+4\Delta\Big(\vec{\mu}_{D(q_1q_2)}\vec{\mu}_{D(q_3c)}+
\vec{\mu}_{D(q_1q_2)}\vec{\mu}_{\bar c}+\vec{ \mu}_{\bar c}\vec{\mu}_{D(q_3c)}\Big)
\nn \\
&&3m_{D(q_1q_2)D(q_3c)\bar c}=
m_{D(q_1q_2)}+m_{D(q_3c)}+m_{\bar c}
\eea
where $\vec{ \mu_D}$ and $\vec{ \mu_{\bar c}}$ are colour-magnetic moments of diquarks and $c$-quark. Diquarks are considered as composite systems of quarks analogous to light nuclei. The magnetic moments are written as sums of quark magnetic moments:
\bea \label{10e}
&&
\vec{\mu}_{D(q_1q_2)}=\frac{\vec{s}_{q_1} }{m_{q_1}}+
\frac{\vec{s}_{q_2}}{m_{q_2}},
\qquad
\vec{\mu}_{D(q_3c)} = \frac{\vec{s}_{q_3} }{m_{q_3}}
+\frac{\vec{s}_c }{m_{c}}\simeq \frac{\vec{s}_{q_3} }{m_{q_3}},\nn\\
&&
\vec{\mu}_c=\frac{\vec{s}_c }{m_{c}}\simeq 0.
\eea
In our estimations we take into account that  $m_{c}>>m_{q} $.

Let us discuss such a scheme for the lowest multiplet of pentaquarks $\bar ccqq'q''$. Estimation of diquark masses is the most problematic issue in the study of diquarks (see for example Ref.~[\citen{santopinto}]). Basing on Refs.~[\citen{?}] and [\citen{QQQQ}] we estimate the masses of diquarks as follows (in MeV units):
\be
\begin{tabular}{ll}
$m_q=330,$ & $m_c=1450$,\\
$m_{S(q_1q_2)}=700$, & $m_{S(q_3c)}=2000$,\\
$m_{A(q_1q_2)}=800$, & $m_{A(q_3c)}=2100$.\\
\end{tabular}
\ee
It gives:
\bea
&&3\mu_{S(q_1q_2)S(q_3c)\bar c}=4150,\quad 3\mu_{A(q_1q_2)A(q_3c)\bar c}=4350,\nn\\
&&3\mu_{S(q_1q_2)A(q_3c)\bar c}=4250,\quad 3\mu_{A(q_1q_2)S(q_3c)\bar c}=4250.
\eea

Here the standard splitting parameter ($\Delta=50$ MeV) is close to that used in the description of splitting of tetraquark states.\cite{QQQQ} It gives the location of the lowest pentaquark multiplet in the mass region 4100-4500 MeV:
\bea\label{ta1}
\begin{tabular}{l|ll}
$ P_{DD_c\bar{c}}^{(L,J^{P})}$& $\qquad mass$  &MeV \\
\hline
$P_{SS_c\bar{c}}^{(0,\frac12^{-})}$ & $3\mu_{SS_c\bar{c}} $ & $\simeq 4150$\\
\hline
$P_{AS_c\bar{c}}^{(1,\frac12^{-})}$ & $3\mu_{AS_c\bar{c}}+2\Delta $ &  $\simeq 4200+100$ =4300\\
$P_{AS_c\bar{c}}^{(1,\frac32^{-})}$& $3\mu_{AS_c\bar{c}}+2\Delta$ & $\simeq 4200+100$ =4300\\
\hline
$P_{SA_c\bar{c}}^{(1,\frac12^{-})}$&$3\mu_{SA_c\bar{c}}$&$\simeq 4250$\\
$P_{SA_c\bar{c}}^{(1,\frac32^{-})}$&$3\mu_{SA_c\bar{c}}$&$\simeq 4250$\\
\hline
$P_{AA_c\bar{c}}^{(0,\frac12^{-})}$ &$3\mu_{AA_c\bar{c}}-4 \Delta$& $\simeq 4350-200$ =4150\\
$P_{AA_c\bar{c}}^{(1,\frac12^{-})}$&$3\mu_{AA_c\bar{c}}-2\Delta$&$\simeq 4350-100$ =4250\\
$P_{AA_c\bar{c}}^{(1,\frac32^{-})}$&$3\mu_{AA_c\bar{c}}-2\Delta$&$\simeq 4350-100$ =4250\\
$P_{AA_c\bar{c}}^{(2,\frac32^{-})}$&$3\mu_{AA_c\bar{c}}+2\Delta$&$\simeq 4350+100$ =4450\\
$P_{AA_c\bar{c}}^{(2,\frac52^{-})}$&$3\mu_{AA_c\bar{c}}+2\Delta$& $\simeq 4350+100$ =4450\\
\end{tabular}
\eea
 In such a scheme all lowest pentaquark states have negative parity. Really in the Glashow formulae the coordinate part of the wave function is hidden in parameters $a$ and $b$. They may be different for standard and exotic hadrons.  Let us emphasize that all mass estimations are qualitative.

Experimental data give us one narrow state $P(4450)$. The consideration of decays gives us reasons to see only one narrow peak: states $\frac12^-$ and $\frac32^-$ can decay in the recombination mode with $s$-wave $J/\Psi$  $p$ states, while the $s$-wave decay of $\frac52^-$ requires the transition $P(4450)\to J/\Psi + N^{\frac32}$. The threshold of $P(4450)\to J/\Psi + N^{\frac32}$ is situated $50Mev$ higher, so the $d$-wave transition $P(4450)\to J/\Psi + p$ is allowed here only. The $d$-wave decay leads to the suppression of width.

Overlapping and mixing of states can also produce broad and narrow resonances. A remarkable feature of the scheme~(\ref{ta1}) is the overlapping of states with equal quantum numbers ($P_{SS_c\bar{c}}^{(\frac12^{-})}$ and $P_{AA_c\bar{c}}^{(\frac12^{-})}$). Such overlapping can produce a strong mixturing of these states. It results in the shift of pole singularities in complex $s$-plane: one resonance becomes narrow with a small width value, the width of the other resonance becomes large.\cite{anis-bugg-sa} The wide resonance can be hardly seen in an observed spectrum.

\subsection{Amplitude poles for multiquark states}
We are based on the hypothesis that amplitude pole singularities exist both for three-quark and five-quark baryons. As an example let us consider the amplitude for $\Delta$-isobar, $\Delta(1240)\to\pi+N$. Due to the decay process this isobar contains both three-quark and five-quark components. In the standard consideration we use only the three-quark component. But if five-quark forces would be strong enough, then the five-quark pole should be situated near the physical region. We suppose that an analogous situation is realized for standard and exotic charmed states.

In terms of the foregoing the pentaquark $P(4450)$ can be considered as a pentaquark with a small admixture of the three-quark state $N^{\frac 52 -}(1675)$ due to annihilation process $(\bar ccuud)^{\frac 52 -}\to(uud)^{\frac 52 -}+(\bar cc)^{0+}$.

\section{New LHCb data}\label{sec2}
In the recent work of LHCb collaboration five narrow peaks were observed which are $\bar ucssu$ system.\cite{LHCbnew}  In the case under consideration the corresponding states can be interpreted as: ${P=\bar u\cdot(cs)\cdot(su)}$ and $P=\bar u\cdot(cu)\cdot(ss)$.

These LHCb data give an opportunity for the estimation of splitting parameter for pentaquarks with open charm. Following the previous section it is a perspective to discuss the estimations for masses of the pentaquark $P(\bar u ussc)$ and versions of classification for observed states.

There are twenty states for pentaquark consisting of $\bar u$, $u$, $s$, $s$ and $c$ quarks in the framework of antiquark-diquark-diquark model. Ten states (see Eq.~\ref{ta1}) for the first choice of diquarks: ${P=\bar u\cdot(cs)\cdot(su)}$ and analogous ten states for the second choice of diquarks: $P=\bar u\cdot(cu)\cdot(ss)$.

In such a model it is reasonable to assume that the highest state formed by an antiquark and two axial diquarks $P_{\bar qAA}^{\frac 52 -}$ refers to $P(3119)$. There are several classification shemes for other states. An example is given below in Eq.~\ref{ta2}. 

\be\label{ta2}
\begin{tabular}{l|l|l|l}
$ P_{\bar uussc}^{J^{P}}$ & mass estimation(MeV) & classification \\
\hline
$P_{\bar uussc}^{\frac12^{-}}$ &3050$\pm$100 & P(3000)\\
\hline
$P_{\bar uussc}^{\frac32^{-}}$ &3200$\pm$100& P(3050), P(3060), P(3090)\\
\hline
$P_{\bar uussc}^{\frac52^{-}}$ &3300$\pm$100&  P(3119)\\
\end{tabular}
\ee
This classification does not exhaust all variants of classifications. This formula is only a qualitative estimation because the mixing of overlapping states is not taken into account. Errorbars of masses in Eq.~(\ref{ta1}) and (\ref{ta2}) are based on the phenomenological description of the systematisation of light meson and baryon states.\cite{ani,b1,b2}

According to the model under consideration two highest states $P^{\frac 52 -}=\bar u\cdot(cs)\cdot(su)$ and $P^{\frac 52 -}=\bar u\cdot(cu)\cdot(ss)$  have close values of masses and hence they are mixing. As a result there is one narrow peak (Eq.~\ref{ta1}) as it was discussed at the end of Section~\ref{sec1}. If among the five narrow peaks two states with $J^P=\frac 52^-$ are seen, it means that the hypothesis of the deutron-like structure of diquarks has failed and formulae (\ref{8}) are not applieble.

Similar to the case of pentaquarks with hidden charm (Sec.~\ref{sec1}) the dumping of the width is the consequence of the large orbital momentum in the $\Xi^+_cK^-$ channel. $s$-wave channel can produce pentaquark with $J=1/2$ only. $J=3/2$ and $J=5/2$ states require $L=2$. Large orbital momentum leads to additional dumping of widths.

LHCb collaboration did not observe any narrow peaks in $\Xi_c^+K^+$ system at 2950 -- 3450 MeV. It seems natural because of existing decay reaction which is located at lower pentaquark masses: $P_{(\bar suusc)}\to\Sigma_c(2455)+(2\pi)^{0++}$. We have to search for the narrow peaks in the system of $\bar s$, $u$, $u$, $s$ and $c$ quarks beginning at the threshold of this reaction -- 2650 MeV.

\section{Conclusion}
We discuss the description of new LHCb data in terms of pentaquark states. Open charm states can be interpreted as three-quark states. \cite{karliner} Predictions of Ref.~[\citen{karliner}] coincide mainly with the predictions of diquark-diquark-antiquark model. It concerns spin-parities and decay modes. Our discussion is motivated by believing that five-quark states exist and corresponding poles should be investigated.

We discuss scenarios for the description of LHCb data.\cite{LHCb,LHCbnew} We use an idea that the corresponding pole singularities can dive into the complex-energy plane due to large widths corresponding to recombination processes. In this situation the narrow states can be observed. Such a state is $P(4450)$ (Sec.~\ref{sec1}). Another mechanism of creating narrow peaks is related to the overlapping of resonance states (Sec.~\ref{sec2}). The calculation of propagators, taking into account loop-diagrams, gives an opportunity to see the effect of specific recombination channels on the mass shifts. Such a procedure was done for tetraquarks. It is done for pentaquark with hidden charm $P(\bar ccqq'q'')$ (to be published). Analogous wave function calculation are to be done for pentaquarks with open charm $P(\bar ssuuc)$ and $P(\bar ussuc)$.

\section*{Acknowledgments}
The authors thank A.K. Likhoded, D.I. Melikhov, A.V. Sarantsev and A.V. Anisovich for useful discussions.

The paper was supported by grant RSF 16-12-10267.

 \end{document}